\documentstyle[12pt,fleqn]{article}
\topmargin - 0.8cm

\catcode`@=11
\@addtoreset{equation}{section}
\catcode`@=12
\mathchardef\SGamma="7100
\begin{document}
\title{\bf Braneworld effective action and origin of inflation}
\author{A.O. Barvinsky$^{a,b\dag}$}
\date{}
\maketitle
\hspace{-8mm}
$^{a}${\em Theory Department, Lebedev Physics Institute,
Leninsky Pr. 53,
Moscow 117924, Russia}\\
$^{b}${\em Physics Department, Ludwig Maximilians University,
Theresienstr. 37, Munich, Germany}

\begin{abstract}
%\xpt
We construct braneworld effective action in two brane Randall-Sundrum
model and show that the radion mode plays the role of a scalar field
localizing essentially nonlocal part of this action. Non-minimal
curvature coupling of this field reflects the violation of
AdS/CFT-correspondence for finite values of brane separation. Under
small detuning of the brane tension from the Randall-Sundrum flat
brane value, the radion mode can play the role of inflaton.
Inflationary dynamics corresponds to branes moving apart in the field
of repelling interbrane inflaton-radion potential and implies the
existence acceleration stage caused by remnant cosmological constant
at late (large brane separation) stages of evolution.  We discuss the
possibility of fixing initial conditions in this model within the
concept of braneworld creation from the tunneling or no-boundary
cosmological state, which formally replaces the conventional moduli
stabilization mechanism.
\end{abstract}
$^{\dag}$e-mail: barvin@td.lpi.ru\\
%\xpt
\section{Introduction}
\hspace{\parindent}
Recently there has been a lot of interest in braneworld scenario
\cite{brane}, inspired, on one hand, by string theory and, on the
other hand, strongly motivated by the attempts to resolve the
hierarchy problem \cite{hierarchy} culminating in 120 orders of
magnitude gap between the Planck scale and the possible value of the
present day cosmological constant \cite{accel}. A simple idea
underlying a potential solution of the hierarchy problem is based on
the possibility to generate exponentially big ratios of energy scales
due to exponential nature of the warped compactification factor in
braneworld models with multi-dimensional AdS bulk \cite{RShier}. This
idea was followed by the observation that in the observable
long-distance approximation the usual 4-dimensional Einstein theory
can be recovered on the 3-brane embedded into AdS 5-dimensional
spacetime \cite{RS,GT}, which fact turned out to be a manifestation
of the widely celebrated by string physicists AdS/CFT-correspondence
\cite{AdS/CFT,SkendSol,Gubser,GKR,HHR1,HHR2}. This made the whole
approach extremely attractive and resulted in numerous attempts,
apparently beginning with \cite{GoldWise}, to stabilize the moduli --
the values of the warped compactification factor at the location of
the visible (Planck) brane -- in the range accounting for the
hierarchy of scales.

Stabilization of moduli, in our opinion, has two important disadvantages.
First, it usually assumes the presence of scalar fields in the bulk which
in certain sense is against the spirit of braneworld scheme -- its
aesthetical purity implies the existence of only the gravitational field
in the bulk. Secondly, stabilization makes braneworld static and, thus,
does not leave a room for inflation and other stages of cosmological
evolution. Question arises whether there exist other, more economical from
the viewpoint of abundance of fields, schemes that could incorporate
cosmological evolution in braneworld picture. The goal of this paper is an
attempt of resolving this question within the two-brane Randall-Sundrum
model in which the role of the inflaton is played by the radion mode --
local field describing the brane separation.

This idea is not entirely new. Brane-separation model of inflaton was
   considered in \cite{DvaliTye} and revisited in \cite{ShiuTye}.
   Recently the model of ekpyrotic scenario \cite{Ekpyr} was put
   forward as an alternative to inflation. Inflation in
   brane-antibrane Universe was considered in \cite{brantibr}. These
   models are based on certain assumptions on the shape of the
   inter-brane potential and qualitatively describe the situation of
   approaching and then colliding branes. Their collision serves
   either as the end of inflation \cite{DvaliTye} (accompanied by
   tachyon mediated brane pairs annihilation and reheating in
   \cite{brantibr}), or as the big bang somehow solving the problems of
   standard scenario \cite{Ekpyr} (see the criticism of the latter in
   \cite{Pyr}).

Qualitatively our model is quite opposite to these suggestions and,
effectively, much simpler because, instead of rather involved
assumptions on the structure of string induced potentials, it relies
   on basic properties of Randall-Sundrum model. Moreover, this model
   brings together a number of issues the combination of which, even
   despite maybe their phenomenological inconsistency, sounds
   promising and deserving further studies. These issues include
   braneworld picture, AdS/CFT-correspondence and the nature of its
   violation implemented in non-minimal curvature coupling of radion,
   radion induced inflation corresponding to {\em diverging} branes
   rather than colliding ones in \cite{DvaliTye,Ekpyr}. They also
   include the issue of initial conditions for inflation actually
   replacing the concept of stabilization and, finally, incorporate
   the phenomenon of cosmological acceleration. The unifying
   framework for all these issues is the formalism of effective
   action induced on branes from the dynamics of gravity in the bulk.

One of the models that served as a motivation for this radion induced
   inflation was the model of the low-energy quantum origin of
   inflationary Universe suggested in \cite{qsi,tunnel}. In this model the
   initial conditions for inflation were generated within the concept of
   tunneling cosmological wavefunction as a probability peak in the
   distribution of the inflaton field $P(\varphi)$. This distribution is
   given by the exponentiated classical Euclidean action of the DeSitter
   instanton $I(\varphi)$ corrected by the loop term which begins with the
   one-loop quantum effective action $\Gamma_{\rm 1-loop}(\varphi)$ of all
   fields inhabiting the instanton background \cite{tunnel},
   $P(\varphi)\simeq\exp[\mp I(\varphi)-\Gamma_{\rm 1-loop}(\varphi)]$
   ($\mp$ sign here is related to no-boundary \cite{HH} and tunneling
   \cite{tun} prescriptions respectively). Applied to the model with
   strong non-minimal curvature coupling of the inflaton
     \begin{eqnarray}
     S[g,\varphi]=\int d^4x \sqrt{g}
     \left\{\frac{m_P^2}{16\pi}R
     -\frac12\xi\varphi^2R +\frac12\varphi\Box\varphi
     -\frac{\lambda\varphi^4}4\right\},
     \,\,\,-\xi=|\xi|\gg 1,     \label{1.1}
     \end{eqnarray}
   this distribution features a sharp probability peak at the value of
     inflaton $\varphi_I$ which corresponds to the inflationary Hubble
     constant $H(\varphi_I)\sim m_P\sqrt{\lambda}/|\xi|$ \cite{qsi}. CMBR
     anisotropy in this model reads in terms of coupling constants
     $\lambda$ and $\xi$ as $\Delta T/T\sim\sqrt{\lambda}/|\xi|$
     \cite{nonmin1,nonmin2}.  Therefore, the energy scale of inflation
   turns out to be much below the Planck scale -- actually even below the
   GUT scale, $H(\varphi_I)\sim 10^{-5}m_P$, so that one can trust
   semiclassical theory. Interestingly, the width of the probability peak
   is also given by the same ratio, $\Delta H/H\sim\sqrt{\lambda}/|\xi|$,
   and the narrow probability peak can be interpreted as the source of
   initial conditions for inflation. The parameters of the matter sector
   of the model can be easily tuned to have sufficiently high e-folding
   number, $N\geq 60$, and this model as a whole can be pretty well
   compatible with observations.

   Big value of $|\xi|$ plays a crucial role in this model. The
   probability peak arises as an artifact of subtle balance between
   the tree-level and one-loop factors in $P(\varphi)$. The one-loop
   effective action can be approximated by the anomalous scaling
   behaviour, $\Gamma_{\rm 1-loop}(\varphi)\sim Z\ln H(\varphi)$,
   where $H(\varphi)$ corresponds to the size -- inverse radius -- of
   the instanton, and $Z$ is the anomalous dimensionality expressible
   in terms of conformal anomaly of quantum matter.  In the slow roll
   approximation due to Higgs effects matter particles acquire masses
   that are much bigger than the curvature scale of the instanton. As
   a result the anomalous scaling $Z$ is dominated by quartic powers
   of these masses and turns out to be quadratic in $\xi$,
   $Z\sim|\xi|^2$. Thus, big $|\xi|$ guarantees that the one-loop
   factor in $P(\varphi)$ strongly suppresses high energy scales and
   generates at intermidiate scales a sharp local peak of the above
   type.

   One of the motivations for the present work was to consider the
   possibility of generating from the braneworld picture such a
   non-minimal inflaton model (\ref{1.1}), thus clarifying the origin
   of inflaton, its non-minimal curvature coupling with big negative
   $\xi$.  As we shall see, the model similar to (\ref{1.1}) can
   really be induced from the two-brane Randall-Sundrum model, the
   role of inflaton being played by the radion mode also
   non-minimally coupled to curvature. The sign and strength of this
   coupling, however, differs from that of (\ref{1.1}) -- the radion
   part of the effective action turns out to be conformal invariant.
   This does not, however, prevent from the inflation scenario.
   Moreover, this scenario is likely to include for late times
   acceleration stage of the Universe. Interestingly, the mechanism
   of generating initial conditions for this braneworld inflation via
   radion field can also be applied here, and it actually replaces
   the stabilization of moduli concept. This is because from the
   viewpoint of formalism this is the same stationarity of the
   effective action requirement as the one in stabilizing the moduli.

   The paper is organized as follows. In Sect.2 we discuss the
   structure of the effective action induced on the brane from bulk
   gravitational dynamics. In particular, we clarify the concept of
   radion as the field in terms of which one can localize essentially
   nonlocal metric action. Regarding the radion mode and its
   dynamical or gauge nature there exists a lot of controversial
   interpretations in current literature. What follows, we hope,
   sheds light on the gauge invariant role of the radion. In Sect. 3
   we present the basics of two-brane Randal-Sundrum model and
   reformulate Garriga-Tanaka's equations for gravitational
   perturbations in this model in the covariant form as an expansion
   in curvature. In Sect. 4 the structure of non-local coefficients
   of this expansion is considered in different limits of brane
   separation and interpreted in terms of AdS/CFT correspondence. The
   low-energy effective action generating these equations is then
   derived in the form containing a non-minimal coupling of the
   radion, which is responsible for strong violation of
   AdS/CFT-correspondence (and zero graviton mode delocalization) for
   small separation of branes. In Sect. 5 this action is cast to the
   form in which both the constant part of brane separation (modulus)
   and its spacetime dependent perturbation are absorbed into one
   field $\varphi$ playing the role of inflaton. This inflaton is
   conformally coupled to Ricci curvature and acquires a nontrivial
   inflaton potential by slightly detuning the brane tension from the
   Randall-Sundrum value characteristic of exactly flat branes. We
   analyze the dynamical behaviour in the Einstein frame and show
   that two branes diverge because of the repelling force of the
   inflaton interbrane potential. Their propagation apart describes
   the inflationary dynamics on the positive tension brane. By
   applying the mechanism of braneworld creation from tunneling state
   we give the estimate for most probable initial conditions of this
   inflationary evolution.  Concluding Sect. 6 contains summarizing
   comments.

   \section{Structure of bulk induced brane effective action}
   \hspace{\parindent}
   In the most general setting the braneworld effective action
   induced from the 5-dimensional bulk looks as follows. We start
   with the theory having the action
     \begin{eqnarray}
     &&S[G,g,\psi]=S_5[G]+\int_{\Sigma} d^4x\sqrt{g}
     \left[L_m(\psi,\nabla\psi)-\sigma\right],         \label{2.0} \\
     &&S_5[G]= \int_{M^5} d^5x\sqrt{ G}
     \left[\,\vphantom{}^5R(G)-2\Lambda_5\right].  \label{2.00}
     \end{eqnarray}
   For clarity of arguments we assume that only
   gravitational field with the metric $G=G_{AB}(x,y)$,
   $A=(\mu,5),\,\mu=0,1,2,3$, lives in the bulk spacetime with
   coordinates $x^A=(x,y),\,x=x^\mu,\,x^5=y$, while matter fields
   $\psi$ with the Lagrangian $L_m(\psi,\partial\psi)$ are confined
   to the brane $\Sigma$ -- 4-dimensional timelike surface embedded
   in the bulk\footnote{For brevity we do not include in the action
   (\ref{2.0}) the surface Gibbons-Hawking term containing the traces
   of extrisic curvatures associated with both sides of the brane
   \cite{ReallGH}.}. This setting can be generalized to the case of
   matter fields propagating in the bulk, but for reasons discussed
   in Introduction we avoid this in what follows. The brane has
   induced metric $g=g_{\mu\nu}(x)$. The bulk and brane parts of the
   action are supplied with the 5-dimensional and 4-dimensional
   cosmological constants. Bulk cosmological constant $\Lambda_5$ is
   negative and, therefore, is capable of generating the AdS
   geometry, while the brane cosmological constant plays the role of
   brane tension $\sigma$ and, depending on the model, can be of
   either sign.

   Full quantum effective action on the brane $S_{\rm eff}[g,\psi]$
   arises as the result of integration over bulk metric subject to the
   boundary condition on the brane -- fixed induced metric which is the
   functional argument of $S_{\rm eff}[g,\psi]$,
     \begin{eqnarray}
     \left.\int DG \exp\{iS[G,g,\psi]\}
     \right|_{\vphantom{I}^4G(\Sigma)=g}=
     \exp\{iS_{\rm eff}[g,\psi]\}.                       \label{2.1}
     \end{eqnarray}
   Note that, with this definition, the matter part of effective action
   coincides with that of the brane action in (\ref{2.0})
     \begin{eqnarray}
     &&S_{\rm eff}[g,\psi]=W[g]+S_m[g,\psi],  \\
     &&S_m[g,\psi]=\int_\Sigma d^4x \sqrt{g}\,L_m(\psi,\nabla\psi),
     \end{eqnarray}
   while all non-trivial dependence on $g$ arising from functional
   integration is contained in $W[g]$. Generically, the calculation of
   this quantity is available only by semiclassical expansion in the
   bulk in powers of$\hbar$. In the tree-level approximation,
     \begin{eqnarray}
     W[g]=S_5[G(g)]-\int_\Sigma d^4x
     \sqrt{g}\sigma+O(\hbar),                  \label{2.3}
     \end{eqnarray}
   $W[g]$ coincides with the 5-dimensional gravitational action on the
   solution of equations of motion {\em in the bulk}, $G=G(g)$, with
   given fixed induced metric on the brane
     \begin{eqnarray}
     &&\frac{\delta S_5[G]}{\delta G_{AB}}=0,  \\
     &&\vphantom{I}^4G_{\mu\nu}(\Sigma)=g_{\mu\nu}.
     \end{eqnarray}

   Given tree-level effective action, one can further apply the variational
   procedure, now with respect to the induced metric $g_{\mu\nu}$, to get the
   4-dimensional equations of motion for the latter
     \begin{eqnarray}
     \frac{\delta S_{\rm eff}}{\delta g_{\mu\nu}}=
     \frac{\delta W}{\delta g_{\mu\nu}}
     +\frac12 T^{\mu\nu}=0,
     \end{eqnarray}
   where $T^{\mu\nu}=(2/\sqrt{g})\delta S_m/\delta g_{\mu\nu}$ is a stress
   tensor of matter fields on the brane. These equations are equivalent to
   the Israel junction conditions
     \begin{eqnarray}
     -\frac1{16\pi G_5}\left[K^{\mu\nu}
     -g^{\mu\nu}K\right]+\frac12 (T^{\mu\nu}-g^{\mu\nu}\sigma)=0
     \end{eqnarray}
   in the conventional treatment of the full system of bulk-brane
   equations of motion (here $\left[K^{\mu\nu}-g^{\mu\nu}K\right]$
   denotes the jump of the extrinsic curvature terms across the
   brane). In our treatment we split the procedure of solving this
   system into two stages. First we solve them in the bulk subject to
   {\em Dirichlet} boundary conditions on the brane and substitute
   the result into bulk action to get {\em off-shell} brane effective
   action. Stationarity of the latter with respect to the 4-dimensional
   metric comprises the remaining set of equations to be solved at
   the second stage. Such splitting might be regarded  as a redundant
   complication, but it allows one to formulate off-shell properties
   of effective braneworld theory and, in particular, analyze it from
   the viewpoint of the AdS/CFT-correspondence
   \cite{Gubser,GKR,HHR1,HHR2}, etc.

   As we see, braneworld effective action induced from the bulk has
   as its arguments only the brane metric and brane matter fields.
   One does not at all meet in such a setting the variables
   describing the embedding of the brane into a bulk. From the
   perspective of 5-dimensional equations of motion this can be
   easily understood, because the dynamical equations for embedding
   variables are automatically enforced in virtue of the bulk
   5-dimensional Einstein equations, Israel junction conditions
   (equivalent, as we have just mentioned, to the stationarity of the
   action with respect to the induced metric) and matter equations of
   motion on the brane\footnote{In view of the 5-dimensional
   diffeomorphism invariance of the full action its variational
   derivative with respect to the embedding variables can be linearly
   expressed in terms of the variational derivatives with respect to
   other fields, which explains this property.}. Another way to look
   at this property is to borrow an old idea from the scope of
   canonical gravity -- the fact that 3-geometry carries information
   about time \cite{BSW}. In braneworld context, this idea implies
   that brane 4-geometry carries the information about the location
   of the brane in the bulk. Indeed, the function (\ref{2.3}) can
   actually be viewed as the Hamilton-Jacobi function of the boundary
   geometry. The only distinction from the canonical theory is that
   the role of boundary is played by the timelike brane, rather than
   spacelike hypersurface, and the role of time is played by the
   fifth (spacelike) coordinate measuring the location of the brane
   (the conventional use of Hamilton-Jacobi equation in cosmology
   in long-wavelength limit \cite{Salopek} was recently extended to
   braneworld context in the number of works including
   \cite{Verlinde,ShToIda}).

   Still, the radion mode is known as an essential ingredient of the
   calculational procedure \cite{GT,ChGR,GKR,HHR2} in braneworld
   theory, and it is worth understanding its role from the viewpoint
   of effective action.  Possible way to recover the radion mode is
   to consider the structure of the effective action $W[g]$
   introduced above. Already in the tree-level approximation this is
   a very complicated nonlocal functional of the metric.  At maximum,
   we know it in the low-energy approximation when the curvature of
   the brane is small compared to the curvature of the AdS bulk
   induced by $\Lambda_5$. The coefficients of the corresponding
   expansion in powers of the curvature are nonlocal. Their
   nonlocality, however, can be of two different types. One type is
   expressible in terms of the massive Green's function of the
   operator $\Box-M^2$ and, therefore, in the low-derivative limit it
   is reducible to the infinite sequence of quasi-local terms, like
     \begin{eqnarray}
     \int d^4x\sqrt{g}R\frac1{\Box-M^2}R\sim
     \int d^4x\sqrt{g}\frac1{M^2}
     \sum\limits_{n}R\left(\frac\Box{M^2}\right)^n\!R.
     \end{eqnarray}
   In the low-energy limit all these terms are suppressed by inverse
   powers of the mass parameter given by the bulk curvature scale,
   $M^2\sim \Lambda_5$, $\Box/M^2\ll1$, and therefore comprise small
   short-distance corrections.  Another type of nonlocality, when
   there is no mass gap parameter like
     \begin{eqnarray}
     \int d^4x\sqrt{g}R\frac1\Box R,\,\,\,
     \int d^4x\sqrt{g}R_{\mu\nu}\frac1\Box R^{\mu\nu},   \label{2.2}
     \end{eqnarray}
   is much stronger in the infrared regime and cannot be neglected.
   Counting the number of degrees of freedom in an essentially
   nonlocal field theory is rather tricky. However, sometimes the
   nonlocal action can be localized in terms of extra fields, which
   makes the local treatment of the theory manageable. For, example,
   structurally the first of nonlocal actions above can be
   reformulated as an expression
     \begin{eqnarray}
     \bar{S}[g,\varphi]\sim\int d^4x\sqrt{g}
     (\varphi R+\varphi\Box\varphi)
     \end{eqnarray}
   containing extra scalar
   field $\varphi$. The variational equation for $\varphi$,
   $\delta\bar{S}/\delta\varphi=0$, yields $\varphi\sim(1/\Box)R$ as
   a solution, which when substituted into $\bar{S}[g,\varphi]$
   renders the original nonlocal structure of
   (\ref{2.2})\footnote{The second of structures (\ref{2.2}) seems to
   require for localization an extra symmetric tensor field.
   Interestingly, though, that within the curvature expansion the
   nonlocal Lagrangian $R_{\mu\nu}(1/\Box)R^{\mu\nu}$ generates the
   variational derivative differing from that of $R(1/\Box)R$ by the
   term proportional to the Einstein tensor \cite{BRath}, so that the
   difference between these two nonlocal Lagrangians can be simulated
   by the Einstein-Hilbert one.}. Thus we suggest that the scalar
   radion mode in the braneworld effective action can be recovered by
   a similar mechanism -- localization of essentially nonlocal
   structures in curvature in terms of the radion field. The fact
   that on shell this field non-locally expresses in terms of
   curvature matches with the known calculations of Ref. \cite{HHR2},
   where the radion mode was given in terms of the conformal part of
   the metric perturbation\footnote{Nonlocal expression for $\xi$ in
   terms of curvature and nonlocalities in the action require
   boundary conditions which depend on particular problem one is
   solving on the 4-dimensional braneworld. For scattering problem
   they are of Feynman chronological nature, for the Cauchy problem
   the non-localities imply retardation. In the no-boundary
   prescription of the cosmological state they can be derived by
   analytic continuation from the Euclidean section of the braneworld
   geometry \cite{HHR1,HHR2}. In what follows we shall not specify
   them explicitly.}. A similar mechanism in 2-dimensional context is
   the localization of the trace anomaly generated Polyakov action
   \cite{Polyakov} in terms of the metric conformal factor. In the
   next two sections we show how such a localization takes place in
   the two-brane Randall-Sundrum scenario.

   \section{Two brane Randall-Sundrum model}
   \hspace{\parindent}
   Here we consider the two-brane Randall-Sundrum model with $Z_2$
   orbifold identification of points on the compactification circle
   of the fifth coordinate \cite{RS}. The action of this model
     \begin{eqnarray}
     &&S_5[G,\psi]=
     \int\limits_{M^5} d^5x\sqrt{G}
     \left[\vphantom{I}^5R(G)-2\Lambda_5\right]  \nonumber\\
     &&\qquad\qquad\qquad+\int\limits_{\Sigma_+} d^4x\sqrt{g_+}
     \left[L_m^+-\sigma_+\right]
     +\int\limits_{\Sigma_-} d^4x\sqrt{g_-}
     \left[L_m^- -\sigma_-\right]                        \label{3.1}
     \end{eqnarray}
   contains the contribution of two branes with two brane tensions
   $\sigma_\pm$ and matter Lagrangians
   $L_m^\pm=L_m(\psi_\pm,\partial\psi_\pm)$. The branes
   $\Sigma_\pm$ are located at antipoidal points of the circle labelled
   by the values of $y$, $y=y_\pm,\,y_+=0,\,|y_-|=d$. $Z_2$-symmetry
   identifies the points on the circle $y$ and $-y$. When the brane
   tensions are opposite in signs and fine tuned in magnitude to the
   values of the negative cosmological constant $\Lambda_5$ and the
   5-dimensional gravitational constant $G_5$ according to the relations
     \begin{eqnarray}
     \Lambda_5=-\frac6{l^2},\,\,\,
     \sigma_+=-\sigma_-=\frac3{4\pi G_5l},   \label{3.2}
     \end{eqnarray}
   then in the absence of matter on branes this model admits the solution
   with the AdS metric in the bulk ($l$ is its curvature radius),
     \begin{eqnarray}
     ds^2=dy^2+e^{-2|y|/l}\eta_{\mu\nu}dx^\mu dx^\nu, \,\,\,
     0=y_+\leq|y|\leq y_-=d,
     \end{eqnarray}
   and with flat induced metric $\eta_{\mu\nu}$ on both branes \cite{RS}.
   The metric on the negative tension brane is rescaled by the value of
   warped compactification factor $\exp(-2d/l)$ providing a possible
   solution for the hierarchy problem \cite{RShier}. With fine tuning
   (\ref{3.2}) this solution exists for arbitrary brane separation $d$
   -- two flat branes stay in equilibrium. Their flatness turns out to
   be the result of a compensation between the bulk cosmological constant
   and brane tensions. This situation includes, in particular, the limit
   of $d\to\infty$ corresponding to only one brane embedded into
   $Z_2$-identified AdS bulk.

   Linearized gravity theory with small matter sources for metric
   perturbations $h_{AB}(x,y)$ on the background of this solution
     \begin{eqnarray}
     ds^2=dy^2+e^{-2|y|/l}\eta_{\mu\nu}dx^\mu dx^\nu
     +h_{AB}dx^Adx^B,                               \label{3.3}
     \end{eqnarray}
   was considered in the series of papers starting with
   \cite{RS,GT,GKR,ChGR}. First, the conclusion on the recovery of the
   4-dimensional Einstein's gravity theory on the brane in the low-energy
   limit was reached in \cite{RS} (graviton zero mode localization) for
   the case of a single brane. It was revised in \cite{GT}, where it was
   shown that in the presence of the second brane one gets the Brans-Dicke
   type theory, rather than the Einstein gravity. The trace of stress
   tensor of matter becomes an additional source of the gravitational
   field due to a nontrivial contribution of the radion mode -- the scalar
   field responsible for the location of brane(s) in the bulk.  This field
   arises as follows.

   By using 5-dimensional diffeomorphism invariance one can transform the
   solution of Einstein equations to the ``Randall-Sundrum
   gauge''\footnote{Strictly speaking this is not a gauge but, rather, a
   corollary of special coordinate conditions and $5A$-components of
   Einstein equations in {\em vacuum} bulk.}
     \begin{eqnarray}
     h_{5\mu}=h_{55}=0,\,\,\,\partial^\mu h_{\mu\nu}=h^\mu_\mu=0.
     \end{eqnarray}
   In this gauge, however, the branes are no longer located at $y=y_\pm$.
   Their embedding equations
     \begin{eqnarray}
     \Sigma_\pm: y=y_\pm+\xi^\pm(x)
     \end{eqnarray}
   involve two scalar functions $\xi^\pm(x)$ \cite{ChGR}, which in the
   linearized theory are of the same order of magnitude as metric
   perturbations, $\xi^\pm(x)\sim h_{\mu\nu}$. They satisfy 4-dimensional
   equations motion \cite{GT}, $\Box\xi^\pm(x)=\pm8\pi G_5 T_\pm(x)/6$,
   in terms of traces of matter stress tensors,
   $T_\pm\equiv\eta^{\mu\nu}T^\pm_{\mu\nu}$, and determine the deviation ]
   of the induced metrics on branes $\bar{h}^\pm_{\mu\nu}$ from the metric
   components $h_{\mu\nu}(x,y_\pm)$ in the Randall-Sundrum gauge
     \begin{eqnarray}
     \bar h_{\mu\nu}^\pm(x)=h_{\mu\nu}(x,y_\pm)
     +l\xi^\pm_{,\mu\nu}(x)
     +\frac2l a^2_\pm\eta_{\mu\nu}\xi^\pm(x)
     +2a^2_\pm\xi^\pm_{(\mu,\nu)}(x).
     \end{eqnarray}
   Here $a_\pm\equiv a(y_\pm)=\exp(-y_\pm/l)$ are the values of warp
   factor on respective branes, and the 4-dimensional vector fields
   $\xi_\mu^\pm(x)$ reflect the ambiguity in two independent
   4-dimensional diffeomorphisms respectively on $\Sigma_+$ and $\Sigma_-$.

   Garriga and Tanaka \cite{GT} wrote down the equations of motion for
   $\bar{h}^\pm_{\mu\nu}(x)$ in the non-covariant form under a certain
   choice of these 4-vector fields. These equations can be easily
   covariantized in terms of Ricci curvatures of brane metrics
   $R^\pm_{\mu\nu}$. For a coupled system of metric perturbations of two
   branes these equations look rather involved and will be presented in
   the coming paper \cite{BRath}. Here we consider a simplified case of
   empty negative tension brane
     \begin{eqnarray}
     T_{\mu\nu}^-=0,\,\,\,\xi^-=0.
     \end{eqnarray}
   It does not experience bending provided we have vanishing Cauchy data
   for the equation for $\xi^-$. Then, the covariantized version of
   Garriga-Tanaka equation for the induced metric on the visible brane
   takes the following form (accompanied by the equation for brane bending
   mode of $\Sigma_+$)
     \begin{eqnarray}
     &&R_{\mu\nu}-\frac12 g_{\mu\nu}R
     =8\pi G(\Box) T_{\mu\nu} +16\pi G_\xi(\Box)
     (\nabla_\mu\nabla_\nu
     -g_{\mu\nu}\Box)\frac1\Box T +O(R^2),  \label{3.6}\\
     &&\Box\xi=\frac{8\pi G_5}6 T.          \label{3.7}
     \end{eqnarray}
   Here all the quantities are related to the positive tension brane
   $\Sigma_+$ and, therefore, we everywhere omit the $+$ label. $O(R^2)$
   indicates that this is a linear order in the
   curvature\footnote{Covariantization of Garriga-Tanaka's equations of
   \cite{GT} in this order of perturbation theory in curvature boils down
   to recovering the Einstein tensor in the left hand side and acquiring
   the $\nabla_\mu\nabla_\nu(1/\Box)T$ term in the right hand side.} and
   other basic quantities $\xi$ and $T_{\mu\nu}$ for which we use
   collective notation $R=(R_{\mu\nu\alpha\beta},\xi,T_{\mu\nu})$.
   Finally, $G(\Box)$ and $G_\xi(\Box)$ are the (generally nonlocal)
   coefficients of the operator nature -- functions of the 4-dimensional
   D'Alambertian. They follow from the Green's function of the linearized
   5-dimensional Eistein equations for metric perturbations in the bulk
   satisfying Neumann type (linearized Israel junction conditions)
   boundary conditions on branes. This Green's function, as known since
   the work of Randall and Sundrum \cite{RS}, includes the contribution of
   zero graviton mode and the tower of massive Kaluza-Klein modes. In the
   low energy limit, $\Box\to 0$, Kaluza-Klein modes give only short
   distance (higher-derivative) corrections, so that these operator
   coefficients
     \begin{eqnarray}
     &&G(\Box)=G_4+O(\Box),               \label{3.4}\\
     &&G_\xi(\Box)=e^{-2d/l}G_4+O(\Box),   \label{3.5}
     \end{eqnarray}
   express in terms of the effective 4-dimensional gravitational constant
     \begin{eqnarray}
     &&G_4=\frac{G_5}l\frac1{1-e^{-2d/l}},\,\,\,
     G_4\to\frac{G_5}l\equiv m_P^{-2}, d\to\infty.
     \end{eqnarray}
   As we see, $G_4$ depends on brane separation $d$ \cite{GT,ChGR} and
   tends to the Randall-Sundrum value in the limit of a single brane,
   $d\to\infty$. In the same limit the Brans-Dicke term containing the
   trace of $T_{\mu\nu}$ vanishes, which corresponds to the recovery
   of Einstein theory in the long distance approximation.

   \section{AdS/CFT correspondence and non-minimal curvature
   coupling of radion}
   \hspace{\parindent}
   As is known, the recovery of low energy Einstein theory for a
   single brane case can be interpreted as a manifestation of the
   AdS/CFT-correspondence in context of braneworld scenario
   \cite{Gubser,GKR,HHR1,HHR2}\footnote{See also \cite{ShToIda} for
   the holographic interpretation of bulk Weyl and curvature squared
   corrections to Einstein equations.}. It is instructive to observe
   this property by considering the structure of
   $O(\Box)$-corrections in (\ref{3.4})-(\ref{3.5}). The Green's
   function of 5-dimensional equations of motion was obtained by
   Garriga-Tanaka in the lowest order approximation in $\Box$,
   leading to (\ref{3.4})-(\ref{3.5}).  Calculations in the
   subleading order of $\Box$-expansion \cite{BRath} show that one
   must distinguish several different energy scales.

   First consider the limit of very large separation between the
   branes and the physical distance on the positive tension brane
   (measured by the magnitude of the quantity $1/\sqrt\Box$) much
   bigger than the curvature radius of the AdS bulk. This corresponds
   to the following two inequalities
     \begin{eqnarray}
     l^2\Box\ll 1,\,\,l^2e^{2d/l}\Box\gg 1.  \label{4.1}
     \end{eqnarray}
   The second inequality actually means that, although the distance
   on the $\Sigma_+$ brane is much greater than $l$, the physical
   distance on $\Sigma_-$, $e^{-d/l}/\sqrt\Box$, lies in the opposite
   range. In this domain the calculation of subleading terms give the
   result \cite{BRath}
     \begin{eqnarray}
     G(\Box)=G_4\left\{1+\frac{l^2\Box}4
     \ln\left(l^2\Box\right)
     +O\left[(l^2\Box)^2\right]\right\}.  \label{4.2}
     \end{eqnarray}
   It contains the logarithmic correction typical of polarization
   effects in local field theory. The coefficients of this
   logarithmic term can apparently be identified with that of a
   specific CFT treated within the $1/N$-expansion -- this has been
   done in many papers on AdS/CFT-correspondence (see, for example,
   \cite{AdS/CFT,SkendSol,HHR1,HHR2,logcoef} regarding the asymptotic
   dependence of $G_5$ and $G_4$ on the number of flavors $N$ in the
   dual CFT theory). Validity of AdS/CFT correspondence in the domain
   (\ref{4.1}) is obvious -- this range implies that the negative
   tension brane is removed far away which, as mentioned above, is
   equivalent to a non-compact single brane case with a continuous
   spectrum of Kaluza-Klein modes. In this case usual arguments of
   AdS/CFT-correspondence apply and we have the logarithmic terms of
   the above type.

   Consider now another range of distances
     \begin{eqnarray}
     l^2\Box\ll 1,\,\,l^2e^{2d/l}\Box\ll 1.     \label{4.3}
     \end{eqnarray}
   This region can be achieved by moving the branes towards one
   another, and it corresponds to the case when the physical distance
   on the negative tension brane, $e^{-d/l}/\sqrt\Box$ becomes
   bigger than the AdS curvature radius. In this range the form
   factor $G(\Box)$ does not have any log-type corrections
   \cite{BRath} and reads
     \begin{eqnarray}
     G(\Box)=G_4\left\{1+\frac{l^2\Box}8\left(3-e^{-2d/l}-\frac{4\,d/l}{1-e^{-2d/l}}\right)
          +O\left[(l^2\Box)^2\right]\right\}.      \label{4.4}
     \end{eqnarray}
   This can obviously be interpreted as a violation of AdS/CFT
   correspondence in the two-brane case. Duality between the
   supergravity theory in the AdS bulk and CFT on its boundary holds
   only in case of a single boundary approaching the infinity of
   5-dimensional AdS spacetime. Thus the AdS/CFT correspondence
   (equivalent in this context to the recovery of Einstein theory on
   the brane) holds only for large brane separation. For small
   separation it gets violated, and the Einstein theory gets deformed
   into the Brans-Dicke type model.

   In what follows we get back to the leading order behaviour of
   nonlocal formfactors (\ref{3.4})-(\ref{3.5}) and substitute them
   into the system of Garriga-Tanaka's equations
   (\ref{3.6})-(\ref{3.7}). Our goal will be to derive the action
   that generates these equations of motion. To do this note that by
   the very definition of $g_{\mu\nu}$ as induced metric on the brane
   it can be directly coupled to the matter stress tensor but {\em
   not to its trace}\footnote{There is a lot of notational confusion
   about this fact in current literature. When the argument of
   effective action is not the induced metric itself but its
   conformally rescaled version, then at the linearized level the
   trace of $T_{\mu\nu}$ can be coupled to perturbations of this
   artificial metric.}. This means that the trace of $T_{\mu\nu}$ has
   to be eradicated from the right hand side of (\ref{3.6}) in favour
   of $\xi$. The radion field $\xi$ for similar reasons cannot be
   coupled to $T$ as well.  Therefore $T$ should also be excluded
   from the right hand side of (\ref{3.7}) -- in terms of the Ricci
   scalar. As a result we have the system of equations

    \begin{eqnarray}
     &&R_{\mu\nu}-\frac12 g_{\mu\nu}R
     =8\pi G_4 T_{\mu\nu} +\frac{2/l}{e^{2d/l}-1}
     (\nabla_\mu\nabla_\nu-g_{\mu\nu}\Box)
     \xi +O(R^2,\Box^2),                    \label{4.5} \\
     &&\Box\xi+\frac l6 R=0.
     \end{eqnarray}
   It is easy to guess the action that directly generates these equations by
   variational procedure with respect to $g_{\mu\nu}$ and $\xi$. It looks
   like
     \begin{eqnarray}
     S_{\rm eff}[g,\xi,\psi]=\int d^4x \sqrt{g}
     \left\{l\,\frac{1-e^{-2d/l}}{16\pi G_5}R
     +\frac{e^{-2d/l}}{8\pi G_5}
     \left(R\xi+\frac3l \xi\Box\xi\right)
     +L_m(\psi,\partial\psi)\right\},               \label{4.6}
     \end{eqnarray}
   where the 4-dimensional gravitational constant $G_4$ is expressed
   in terms of the fundamental 5-dimensional one, $G_5$, in order to
   have explicit dependence on the brane separation $d$. As we see,
   the radion mode $\xi$ is non-minimally coupled to metric via Ricci
   curvature, and there is no direct coupling of $\xi$ to matter.
   Quite similarly, matter is coupled to the brane metric via the
   matter Lagrangian which generates $T_{\mu\nu}$ by usual
   variational procedure. Thus, the low-energy effective action turns
   out to be a local functional in terms of metric and the radion
   mode. However, the exclusion of $\xi$ in terms of the curvature
   scalar, $\xi=-(l/6)(1\Box)R$, leads, along the lines outlined in
   the previous section, to essentially nonlocal metric part
   (\ref{2.3}) of the whole effective action
     \begin{eqnarray}
     W[g]=\frac l{16\pi G_5}\int d^4x \sqrt{g}
     \left\{(1-e^{-2d/l})R-\frac16 e^{-2d/l}
     R\frac1\Box R+O(R^3)\right\}.                         \label{4.7}
     \end{eqnarray}
   This confirms the mechanism suggested in Sect.2.

   \section{Radion mode as an inflaton field}
   \hspace{\parindent}
   Let us get back to the local action (\ref{4.6}). It was obtained in
   the approximation quadratic in curvature and radion mode. However,
   its structure clearly indicates that it can be regarded as an expansion
   in $\xi$ of the following action
     \begin{eqnarray}
     S_{\rm eff}[g,\varphi]=\int d^4x \sqrt{g}
     \left\{\frac{m_P^2}{16\pi}R+\frac12
     \left(\varphi\Box\varphi
     -\frac16 R\varphi^2\right)\right\}               \label{5.1}
     \end{eqnarray}
   with the new scalar field $\varphi$ exponentially related to the original
   small radion mode
     \begin{eqnarray}
     \varphi(x)=\sqrt{\frac3{4\pi}}m_P
     \exp\left[-\frac{d+\xi(x)}l\right],
     \,\,\,m_P^2=\frac l{G_5}.               \label{5.2}
     \end{eqnarray}

   The fact that the initial separation of branes $d$ (moduli)
   combines with the spacetime dependent perturbation of this
   separation $\xi(x)$ into the variable $d+\xi(x)$ is geometrically
   very natural, because this variable describes a movement and local
   bending of the brane relative to the (invisible) negative tension
   brane. In this parametrization the dependence of effective action
   on brane separation is entirely absorbed into the new field
   $\varphi(x)$. The limit of infinitely big separation between
   branes  corresponds to vanishing $\varphi$, $\varphi(x)\to 0$,
   $d+\xi(x)\to\infty$, and a complete recovery of Einstein theory.
   For intermidiate range of brane separation the field $\varphi$
   becomes dynamically important. It is non-minimally coupled to
   curvature like in (\ref{1.1}), but the coupling constant,
   $\xi=1/6$, is positive and renders the $\varphi$-dependent part of
   the action to be conformally invariant\footnote{This form of the
   action was also obtained in Ref.\cite{GoldWise1} by Kaluza-Klein
   reduction, without revealing, however, its conformal nature. I am
   grateful to H.Tye for drawing my attention to this work.}. The
   kinetic term of $\varphi$ has a good positive sign, so that there
   is no instability associated with the positive tension
   brane\footnote{Instability of two brane system is associated with
   the negative tension brane, on which the corresponding brane
   bending mode has a ghost nature \cite{BRath,Pilo}.}.

   As a whole the action (\ref{5.1}) is not conformal invariant
   because of the first -- Einstein-Hilbert -- term, which gives hope
   that the radion field $\varphi$ can play the role of inflaton.
   However, this field does not, thus far, have any potential that
   could have induced the slow roll scenario. A possible mechanism to
   get a radion potential is to make a small detuning \cite{Cline} of
   the brane tension $\sigma=\sigma_+$ from the Randall-Sundrum value
   (\ref{3.2}). Let us denote excessive part of the brane tension on
   the positive tension brane by $\sigma_e$, $\sigma=3/4\pi
   G_5l+\sigma_e$. This additional positive tension can be viewed as
   a result of the replacement of the original matter Lagrangian by
   the one containing small 4-dimensional cosmological term
     \begin{eqnarray}
     L_m(\psi,\nabla\psi)\to
     L_m(\psi,\nabla\psi)-\sigma_e.
     \end{eqnarray}
   The inclusion of this term into the metric-radion part of the action
   gives rise to the action
     \begin{eqnarray}
     S_{\rm eff}[g,\varphi]=\int d^4x \sqrt{g}
     \left\{\left(\frac{m_P^2}{16\pi}
     -\frac1{12} \varphi^2\right)R
     +\frac12\varphi\Box\varphi-\sigma_e \right\},   \label{5.10}
     \end{eqnarray}
   which still does not have a good inflaton potential as a growing
   function of $\varphi$. Note, however, that $\varphi$ is
   non-minimally coupled to curvature, which makes inflationary
   implications of this model less trivial
   \cite{nonmin1,nonmin2,Linde,efeq}. To analyse them we go over to
   the Einstein frame of the conformally related metric
   $\bar{g}_{\mu\nu}$ and new scalar field,
   $(g_{\mu\nu},\varphi)\to(\bar{g}_{\mu\nu},\phi)$ \cite{renorm,efeqmy},
     \begin{eqnarray}
     &&g_{\mu\nu}=\cosh^2\left(\sqrt{4\pi/3}\,
     \frac\phi{m_P}\right)\bar{g}_{\mu\nu},             \\
     &&\varphi=\sqrt{\frac3{4\pi}}m_P\tanh\left(\sqrt{4\pi/3}\,
     \frac\phi{m_P}\right).
     \end{eqnarray}
   In this parametrization the range of the original field $\varphi$,
   $|\varphi|\leq\sqrt{3/4\pi}m_P$, in which the effective
   $\varphi$-dependent gravitational constant of the action (\ref{5.10})
     \begin{eqnarray}
     \frac1{16\pi G_4(\varphi)}=\frac{m_P^2}{16\pi}
     -\frac1{12}\varphi^2
     \end{eqnarray}
   is positive, is mapped onto the infinite range of $|\phi|<\infty$. This
   guarantees that we are considering a stable phase of the theory free of
   ghosts\footnote{Graviton modes have a ghost nature outside of this range
   which, however, corresponds to $d+\xi(x)<0$, and thus can be ruled out on
   physical ground, because $d+\xi(x)=0$ implies the limit when the branes
   locally touch one another at the point $x$.}.

   The action in the Einstein frame (barred quantities are defined relative
   to the barred metric $\bar{g}_{\mu\nu}$)
     \begin{eqnarray}
     &&\bar{S}_{\rm eff}[\bar{g},\phi]=\int d^4x \sqrt{\bar{g}}
     \left\{\frac{m_P^2}{16\pi}\bar{R}
     +\frac12\phi\bar\Box\phi-\bar{V}(\phi) \right\},   \\
     &&\bar{V}(\phi)=\sigma_e\cosh^4\left(\sqrt{4\pi/3}\,
     \frac{\phi}{m_P}\right),
     \end{eqnarray}
   has a minimally coupled field $\phi$ and the potential {\em
   monotonically growing} from its minimum value to infinity. Minimum
   of the potential corresponds to an infinite separation of branes,
   while the infinity of $\bar{V}(\phi)$  describes the limit of
   coincident branes with $\varphi=\sqrt{3/4\pi}m_P$, $d+\xi=0$. This
   means that the branes are {\em repelling} in our setting: the
   dynamical roll of the radion field down the potential well leads
   to the diverging branes. The visible brane with slightly detuned
   tension is curved and in the slow-roll regime has quasi-DeSitter
   geometry embedded into an AdS 5-dimensional bulk. For large
   initial values of $\phi$ (small brane separation) this potential
   might be too steep to maintain the slow roll regime, but it can
   realize the power law inflation scenario, while for small $\phi$
   ($\phi\ll\sqrt{3/4\pi}m_P$, big separation) both conditions of the
   slow roll approximation are satisfied perfectly well. Thus, the
   radion mode can really be a candidate for the inflaton field
   generating inflation.

   Note that in our model no stabilization mechanism is involved for
   fixing the value of the radion moduli. Two branes initially
   located at some separation start diverging from one another and
   induce inflation on the positive tension brane. The parameters of
   this inflation certainly depend on the initial conditions. To
   determine them one  can invoke the ideas of quantum cosmology --
   the creation of the braneworld described by the gravitational
   instanton either within the no-boundary or tunneling proposals for
   the cosmological wavefunction. In the tree-level approximation the
   brane-world creation was considered in \cite{GarSas}.  It was
   shown that for large brane separation of concentric DeSitter
   branes the probability amplitude is dominated by the contribution
   of the 4-dimensional instanton action of the positive tension
   brane. In our model this is given in terms of the radion potential
   as
     \begin{eqnarray}
     I(\phi)=-\frac{3m_P^4}{8\bar{V}(\phi)}.
     \end{eqnarray}
   Modified by the contribution of one-loop effective action (in the
   approximation of the overall scaling behaviour with
   $H(\phi)\sim\cosh(\sqrt{4\pi/3}\phi/m_P)$), the
   distribution function of the initial value of the radion field
   (cf. Introduction)
     \begin{eqnarray}
     P(\phi)\sim\exp\left(\pm I(\phi)-Z\ln\cosh(\sqrt{4\pi/3}\,
     \phi/m_P)\right)
     \end{eqnarray}
   can have a probability maximum for the tunneling case (sign + in
   the exponential) corresponding to the effective Hubble constant
   $H\sim(\sigma_e/Z)^{1/4}$, provided the parameter $\sigma_e$
   satisfies the upper bound $\sigma_e\leq 3m_P^4/2Z$. For very big
   values of the anomalous scaling parameter $Z$ of quantum matter
   inhabiting the brane these estimates can bring us to reasonable
   parameters of the inflationary scenario\footnote{For $SU(N)$
   conformal field theory $Z$ -- conformal anomaly integrated over
   the instanton 4-volume -- is proportional to $N^2$
   \cite{HHR1,HHR2} and therefore can be big within
   $1/N$-expansion.}. In fact, the extremum of the radion probability
   distribution which supplies us with initial conditions replaces
   the equations for stabilizing braneworld moduli (because in both
   cases the effective action is being analyzed for its extremum).
   However, in contrast with the stabilization scheme, we fix only
   initial conditions, the further evolution of the radion is not
   frozen and mimicks inflation.

   Important fact is that the potential does not vanish at its
   minimum, so that inflation never ends. This might be bad from the
   viewpoint of classical inflation scenario. However, in view of the
   recent discovery of acceleration of the Universe \cite{accel},
   this property can be interpreted as the fact that the radion
   induced inflation scenario also includes the mechanism for the
   present day acceleration of the Universe\footnote{The author is
   grateful to Lev Kofman for drawing attention to this fact.}.

   \section{Conclusions}
   \hspace{\parindent}
   Thus we have shown that the radion mode in braneworld dynamics
   arises as a scalar field localizing essentially nonlocal part of
   braneworld metric effective action. For a positive tension brane,
   this mode has good dynamical properties of its kinetic term and,
   in the case of a positive detuning of the brane tension on this
   brane in the two brane Randall-Sundrum model, it can generate
   inflationary scenario via non-minimal coupling to the curvature.
   Inflationary evolution arises as a manifestation of the fact that
   two branes move apart from one another due to the repelling force
   mediated by this non-minimal coupling. Thus, this scenario is
   quite opposite to the picture of ekpyrotic Universe \cite{Ekpyr}
   and brane separation models of \cite{DvaliTye,ShiuTye} and
   \cite{brantibr}. On the other hand, the suggested model does not
   incorporate any stabilization mechanisms for brane moduli. Rather,
   the formalism quite similar to that of stabilization leads to the
   determination of the initial conditions for radion dynamics via
   extremizing the (loop corrected) probability distribution of the
   radion field.

   Full analyses of the inflationary scenario in this model goes
   beyond the scope of this paper.  Phenomenologically it might be
   too naive to accommodate the observational bounds on inflation and
   acceleration of our Universe. One might, for example, expect
   serious difficulties with the preheating mechanism of transition
   from the inflationary stage to the radiation dominated one.
   Moreover, it can hardly explain the magnitude of the present day
   acceleration, even though this model contains a natural mechanism
   of this acceleration due to remnant cosmological constant that
   survives the limit of large brane separation (corresponding to
   present time). However, as a whole it looks very promising because
   it naturally brings together the whole set of intriguing issues --
   fundamental origin of inflaton as a mechanism of violation of
   AdS/CFT-correspondence, issue of initial conditions for inflation
   replacing the idea of stabilization mechanisms and cosmological
   acceleration. It is quite possible that the synthesis of this
   model with subtle mechanisms for generating bulk cosmological
   constants, brane tensions \cite{Tye} and mechanisms of their
   detuning \cite{Cline} can result in phenomenologically viable
   model approaching the solution of cosmological hierarchy problem.

   \section*{Acknowledgements}
   \hspace{\parindent}
   The author benefitted from helpful discussions with
   C.Armendariz-Picon, L.Kofman, V.Rubakov, I.Sachs, S.Solodukhin and
   H.Tye and is also grateful to Andreas Rathke for a hard task of
   checking many calculations in this paper. He is grateful for
   hospitality of the Physics Department of LMU, Munich, where a
   major part of this work has been done. This work was supported by
   the Russian Foundation for Basic Research under the grant No
   99-01-00980.

   \end{document}